\documentclass[conference]{IEEEtran}

\AtBeginDocument{%
 \providecommand\BibTeX{{%
   \normalfont B\kern-0.5em{\scshape i\kern-0.25em b}\kern-0.8em\TeX}}}

\usepackage{colortbl}
\usepackage{xspace}
\newcommand{\ie}{\textit{i.e., \xspace}}
\newcommand{\eg}{\textit{e.g., \xspace}}
\newcommand{\etal}{\textit{et al. \xspace}}

\usepackage{array}
\newcolumntype{L}{>{\arraybackslash}m{16cm}}
\newcolumntype{C}[1]{>{\centering\let\newline\\arraybackslash\hspace{0pt}}m{#1}}
\newcolumntype{R}[1]{>{\raggedleft\let\newline\\arraybackslash\hspace{0pt}}m{#1}}
\usepackage[numbered]{bookmark}
\newcommand{\ballnumber}[1]{\tikz[baseline=(myanchor.base)] \node[circle,fill=.,inner sep=1pt] (myanchor) {\color{-.}\bfseries\footnotesize #1};}
\usepackage[many]{tcolorbox}
\usepackage{graphicx}
\usepackage[switch]{lineno}
\usepackage{fontawesome}
\usepackage{lineno}
\usepackage{xcolor}
\usepackage{stfloats}
\usepackage[justification=centering]{caption}
\usepackage{dirtytalk}
\usepackage{amsmath}
\usepackage{pgfplots, pgfplotstable}
\usetikzlibrary{patterns}
\usepackage{supertabular} 
\usepackage[flushleft]{threeparttable}
\usepackage{supertabular}

\usepackage[english]{babel}
\usepackage{textcomp}
\usepackage{mathtools}
\usepackage{ltxtable}
\usepackage{algorithm}
\usepackage{algpseudocode}
\usepackage{textcomp}

\usepackage{pgf-pie}
\definecolor{wedge1}{RGB}{ 190  30  46}
\definecolor{wedge2}{RGB}{ 240  65  54}
\definecolor{wedge3}{RGB}{ 241  90  43}
\definecolor{wedge4}{RGB}{ 247 148  30}
\definecolor{wedge5}{RGB}{  43  56 144}
\definecolor{wedge6}{RGB}{  28 117 188}
\definecolor{wedge7}{RGB}{  40 170 225}
\definecolor{wedge8}{RGB}{ 119 179 225}
\definecolor{wedge9}{RGB}{ 181 212 239}
\definecolor{wedge10}{RGB}{  0 104  56}
\definecolor{wedge11}{RGB}{  0 148  69}
\definecolor{wedge12}{RGB}{ 57 181  74}
\definecolor{wedge13}{RGB}{141 199  63}
\definecolor{wedge14}{RGB}{215 244  34}
\definecolor{wedge15}{RGB}{249 237  50}
\definecolor{wedge16}{RGB}{248 241 148}
\definecolor{wedge17}{RGB}{242 245 205}
\definecolor{wedge18}{RGB}{123  82  49}
\definecolor{wedge19}{RGB}{104  73 158}
\definecolor{wedge20}{RGB}{102  45 145}
\definecolor{wedge21}{RGB}{148 149 151}
\definecolor{wedge22}{RGB}{ 204 50 153}
\definecolor{wedge23}{RGB}{ 79 47 79}
\definecolor{wedge24}{RGB}{ 173 234 234}
\definecolor{wedge25}{RGB}{ 216 191 216}
\definecolor{wedge26}{RGB}{  43  56 144}
\definecolor{wedge27}{RGB}{  40 170 225}
\definecolor{wedge28}{RGB}{ 119 179 225}
\definecolor{wedge29}{RGB}{ 181 212 239}
\definecolor{wedge30}{RGB}{  0 104  56}
\definecolor{wedge31}{RGB}{  0 148  69}
\definecolor{wedge32}{RGB}{ 57 181  74}

\usetikzlibrary{chains}

\pgfplotstableread[col sep=comma,]{data.csv}\datatable

\pgfplotstablecreatecol[
    create col/expr={( \thisrow{B} ) / ( \thisrow{B} + \thisrow{C} + \thisrow{D} + \thisrow{E} + \thisrow{F} )}] {B_fraction} \datatable
\pgfplotstablecreatecol[
    create col/expr={( \thisrow{C} ) / ( \thisrow{B} + \thisrow{C} + \thisrow{D} + \thisrow{E} + \thisrow{F} )}] {C_fraction} \datatable
\pgfplotstablecreatecol[
    create col/expr={( \thisrow{D} ) / ( \thisrow{B} + \thisrow{C} + \thisrow{D} + \thisrow{E} + \thisrow{F} )}] {D_fraction} \datatable
\pgfplotstablecreatecol[
    create col/expr={( \thisrow{E} ) / ( \thisrow{B} + \thisrow{C} + \thisrow{D} + \thisrow{E} + \thisrow{F} )}] {E_fraction} \datatable
\pgfplotstablecreatecol[
    create col/expr={( \thisrow{F} ) / ( \thisrow{B} + \thisrow{C} + \thisrow{D} + \thisrow{E} + \thisrow{F} )}] {F_fraction} \datatable

\pgfmathsetmacro\startAngle{90-3.6/2}
\pgfmathsetmacro\radius{+5}
\pgfmathsetmacro\maxLeg{+12}
\pgfmathsetmacro\legBound{+60}
\pgfmathsetmacro\legSpacing{2*\legBound/(\maxLeg-1)}
  \pgfplotstableread[row sep=\\,col sep=&]{
        interval & carT & sd\\
     }\mydata

\usepackage{verbatim}
\pgfplotsset{compat=1.15}
\usetikzlibrary{patterns,}

\usepackage{graphicx}
\usepackage{tabularx,booktabs}
\usepackage{array}
\usepackage{bbding}
\usepackage{pifont}
\usepackage{wasysym}
\usepackage{caption}
\usepackage{dirtytalk}
\usepackage{subcaption}
\usepackage{tabularx}
\usepackage{sfmath}
\usepackage{array, booktabs, makecell}
\usepackage{color}
\usepackage{csquotes}
\usepackage{hyperref}
\usepackage{comment}
\usepackage{tikz}
\usepackage{balance}
\usepackage{listings}
\usepackage{booktabs} 
\usepackage{soul} 
\usetikzlibrary{fit} 
\usetikzlibrary{positioning}
\usetikzlibrary{arrows}
\usetikzlibrary{shapes.multipart}
\usepackage{adjustbox}
\usepackage{float}
\usepackage{rotating}
\usepackage{nth}
\DeclareCaptionType{TextBox}
\usepackage{pstricks}
\usepackage{placeins}
\usepackage{afterpage}
\usepackage{multirow} 
\usepackage{textcomp}
\usepackage{multicol}
\usepackage{varwidth} 

\definecolor{findOptimalPartition}{HTML}{D7191C}
\definecolor{storeClusterComponent}{HTML}{FDAE61}
\definecolor{dbscan}{HTML}{ABDDA4}
\definecolor{constructCluster}{HTML}{2B83BA}

\newcommand{\todo}[1]{\textcolor{red}{{\it [Added]}}}
\definecolor{main}{HTML}{5989cf}    
\definecolor{sub}{HTML}{cde4ff}     
\tcbset{
    sharp corners,
    colback = white,
    before skip = 0.2cm,    
    after skip = 0.5cm      
}   
\newtcolorbox{boxE}{
    enhanced, 
    boxrule = 0pt, 
    borderline = {0.75pt}{0pt}{main}, 
    borderline = {0.75pt}{2pt}{sub} 
}
\newtcolorbox{boxK}{
    sharpish corners, 
    boxrule = 0pt,
    toprule = 4.5pt, 
    enhanced,
    fuzzy shadow = {0pt}{-2pt}{-0.5pt}{0.5pt}{black!35} 
}
\usepackage{listings}

\definecolor{javared}{rgb}{0.6,0,0} 
\definecolor{javagreen}{rgb}{0.25,0.5,0.35} 
\definecolor{javapurple}{rgb}{0.5,0,0.35} 
\definecolor{javadocblue}{rgb}{0.25,0.35,0.75} 
 
\definecolor{light-red}{rgb}{1,0.92,0.91}
\definecolor{light-green}{rgb}{0.9,1,0.93}

\author{
\IEEEauthorblockN{Eman Abdullah AlOmar\IEEEauthorrefmark{1}, 
Catherine DeMario\IEEEauthorrefmark{1},
Roger Shagawat\IEEEauthorrefmark{1},
Brandon Kreiser\IEEEauthorrefmark{1}}
\IEEEauthorblockA{\IEEEauthorrefmark{1}Stevens Institute of Technology, Hoboken, New Jersey, USA\\
\{ealomar,cdemario,rshagawa,bkreiser\}@stevens.edu\\
}} 

\newcommand{\toolname}{\textsc{LeakageDetector}\xspace}

\begin{document}


\title{LeakageDetector: An Open Source Data Leakage Analysis Tool in Machine Learning Pipelines}


\maketitle
\begin{abstract}
Code quality is of paramount importance in all types of software development settings. Our work seeks to enable Machine Learning (ML) engineers to write better code by helping them find and fix instances of Data Leakage in their models. Data Leakage often results from bad practices in writing ML code. As a result, the model effectively \say{memorizes} the data on which it trains, leading to an overly optimistic estimate of the model performance and an inability to make generalized predictions. ML developers must carefully separate their data into training, evaluation, and test sets to avoid introducing Data Leakage into their code. Training data should be used to train the model, evaluation data should be used to repeatedly confirm a model's accuracy, and test data should be used only once to determine the accuracy of a production-ready model. In this paper, we develop \toolname, a Python plugin for the PyCharm IDE that identifies instances of Data Leakage in ML code and provides suggestions on how to remove the leakage.  The plugin and its source code are publicly available on GitHub at \url{https://github.com/SE4AIResearch/DataLeakage\_Fall2023}. The demonstration video can
be found on YouTube: \url{https://youtu.be/yXj3wihSaIU}.

\end{abstract}




\section{Introduction}
\label{Section:Introduction} 
Data Leakage is one of the critical issues in writing Machine Learning (ML) that can significantly affect the performance and reliability of models \cite{burkov2020machine,yang2022data,kaufman2012leakage,apicella2024don,drobnjakovic2024abstract,liu2023program,epperson2022strategies,kery2018story,koenzen2020code,shimakawa2024prevention,babu2024improving}. It occurs when information outside the training dataset is inadvertently used to create the model, leading to overly optimistic performance and poor generalization to new data. Data Leakage often arises from bad practices in data handling and code implementation, making it a pervasive challenge in the field \cite{nahar2022collaboration,hulten2018building,amershi2019software,chorev2022deepchecks,kaufman2012leakage,subotic2022static}. Previous studies have shown that Data Leakage is a prevalent bad practice, particularly in educational settings \cite{michalski2013machine,burkov2020machine}. 

Recently, Yang \etal \cite{yang2022data} have proposed a static analysis approach to detect Data Leakage, that is a command-line tool that can detect common forms of Data Leakage (\ie Overlap, Multi-test, and Preprocessing) in data science code. However, despite efforts to engineer an automated detection for Data Leakage, the tool is designed for research purposes, such as enabling large-scale empirical studies. The tool could benefit from some aspects, such as usability, user engagement, and leakage correction, which enhances practitioners' adoption.

To cope with these challenges, this paper aims to support developers with the identification of Data Leakage by designing \toolname, a PyCharm IDE plugin that detects and proposes quick fixes for Data Leakage instances.

Our plugin enhances the functionality of the leakage static analysis tool \cite{yang2022data} by integrating directly with the PyCharm IDE, making it more user-friendly and accessible. Unlike the original tool, which requires command-line proficiency and manual HTML file review, our plugin operates within the IDE with a single click, providing clear, plain-language descriptions of different types of leakage. It simplifies deployment, scales well with the user base due to PyCharm's popularity, and offers customization features such as quick fixes and \say{TODO} notes.

\begin{figure*}[htbp]
  \centering
  \includegraphics[width=1.0\textwidth]{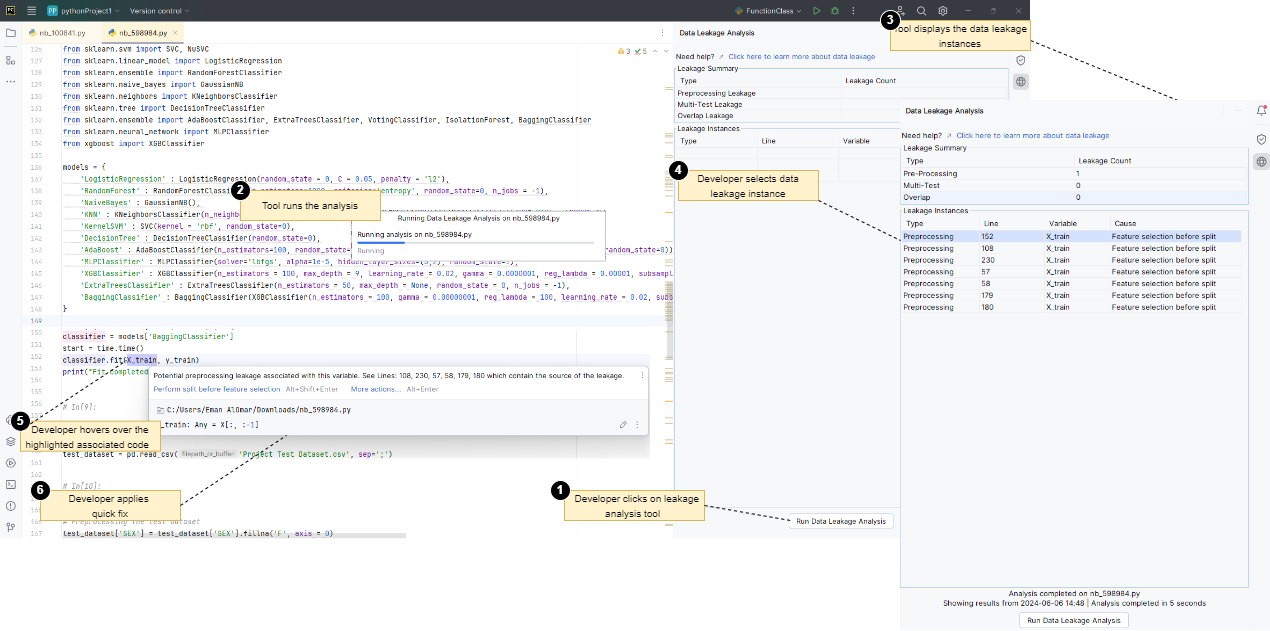}
  \caption{\toolname in action, showing the identified Data Leakage instances.}
  \label{fig:example}
  \vspace{-0.3cm}
 \end{figure*}

As an illustrative example, Figure \ref{fig:example}, shows a step-by-step scenario to automatically detect and fix the Data Leakage instance. Once the user opens a Python file within PyCharm, and opens the Data Leakage Analysis Tool Window,  the user clicks on the \say{Run Data Leakage Analysis} button to run the leakage analysis tool \ballnumber{1}. When leakage analysis is complete \ballnumber{2} and code segments associated with Data Leakage are highlighted, the same code segments are shown in the Tool Window \ballnumber{3}. The user selects the Data Leakage instance \ballnumber{4} and can hover over the highlighted code associated \ballnumber{5} and apply the quick fix \ballnumber{6}.

The preliminary evaluation of our tool shows its ability to detect and suggest quick fixes for the selected Data Leakage instances. The survey responses
are promising, as the majority of participants are satisfied with the
recommendations of \toolname. 

The remainder of this paper is organized as follows: Section \ref{Section:methodology} outlines our tool design. Section \ref{Section:Result} discusses our preliminary evaluation, while the research discussion is addressed in Section \ref{Section:Implication}. Section \ref{Section:RelatedWork} reviews
the existing studies related to the Data Leakage,
before concluding with Section \ref{Section:Conclusion}.


\section{Study Design} 
\label{Section:methodology}

\subsection{\toolname Architecture}

\begin{figure}[htbp]
  \centering
  \includegraphics[width=1.1\columnwidth]{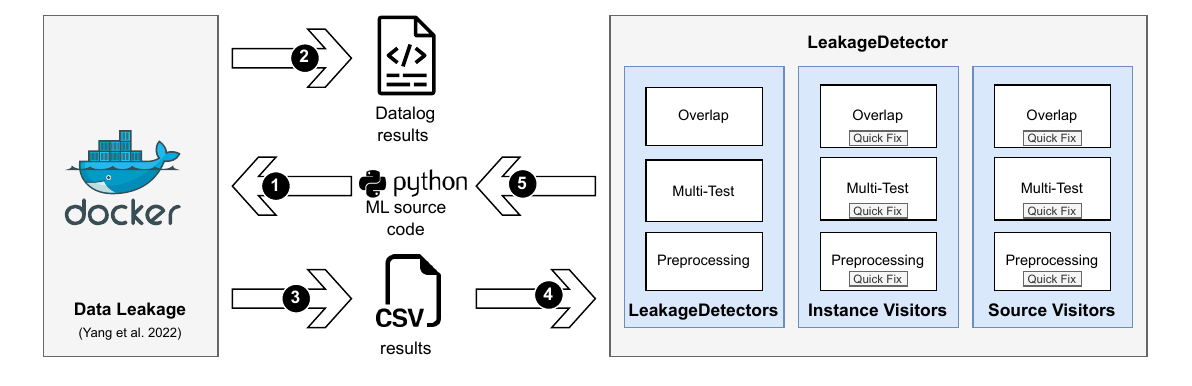}
  \caption{High-level architecture of \toolname.}
  \label{fig:arch}
  \vspace{-0.3cm}
 \end{figure}

\toolname is implemented as an open-source Python plugin for PyCharm IDE. A high-level overview
of the architecture of \toolname is depicted in Figure \ref{fig:arch}. While the user works on the PyCharm IDE with the plugin installed, the user can navigate to the Data Leakage Tool Window on the screen's right side. Located at the bottom of the Tool Window, there is a button labeled \say{Run Data Leakage Analysis}, which invokes the following process: The plugin checks whether the Docker image of the Leakage Analysis tool \cite{yang2022data} is downloaded on the user’s machine. Once the image is on the user’s machine, the plugin connects to the Docker engine, creates a container from the image, and uses that container to analyze the currently open Python file. Then, the output from the tool is placed in a temporary folder on the user's machine. Our plugin classes contain functionality that examines these files to determine whether leakage is present in the user's Python file. The plugin’s inspection visitors will use the information from the leakage detectors to determine where to render inspections in the Python file. If there is a leakage, the plugin will perform inspections (\ie highlight/underline) of the code involved in the leakage. The user can apply a quick fix to a piece of code involved in the leakage. The plugin keeps track of the lines affected by the quick fix in a file on the user’s machine.


Our plugin builds on the functionality of the leakage static analysis tool \cite{yang2022data} in a number of key ways:
\begin{itemize}
    \item 	\textbf{Functionality.} Compared to the Data Leakage static analysis tool, our plugin can be run within the IDE. Originally, if a user wanted to run the tool, they would have to run a command line and then open an HTML file with the output. Having the tool already in the IDE makes the tool much easier for anyone to use. 
\item \textbf{User Interaction.} The Data Leakage static analysis tool requires users to be proficient in using command-line commands and SWIG to run the tool. This can be a problem for those who are new to coding. Once installed in PyCharm and alongside a running Docker, our plugin requires just a push of a button to run. Our plugin easily displays plain-language descriptions of the different types of leakage in the IDE for users to see and understand. 

\item 	\textbf{Integration.} The plugin is already integrated with PyCharm compared to the original tool, which has no integration with any IDE. 
\item 	\textbf{Ease of Deployment.} The plugin does require some extra steps, such as installing Docker and installing the plugin in PyCharm; compared to the tool, it is just a couple of command lines. Yet, once all the prerequisites to the plugin are installed, the plugin is easier to use, as the plugin can run with a click of a button. 
\item	\textbf{Scalability.} Both the plugin and the tool have a similar scale in terms of the amount of code they can detect leakage on (\ie 1 file). Regarding audience reach, we believe that the plugin would be more accessible to a greater audience as PyCharm is a popular IDE machine-learning tool that developers like to use. 
\item	\textbf{Customization.} The plugin and the tool both have the ability to identify leakages and enable users to navigate to the specific line where the error occurs. the plugin has the added benefit of allowing users to do a quick fix (see Subsection \ref{sec:quick-fix}), which would change the code and add a \say{TODO} for the users explain what needs to be fixed to improve their code. 
\item	\textbf{Feedback Mechanism.} Both the plugin and the tool show similar results (\eg type of Data Leakage, line number) in a table. Still, the tool shows the table at the top of the HTML result, and when a user scrolls down, the result table does not follow. On the other hand, the plugin would have the table displayed in the tool itself, which would be on the right side of the IDE. Also, the plugin will always be on the right side if the user scrolls down through their code. 
\end{itemize}

We believe that our plugin complements the command-line-based tool \cite{yang2022data}, providing users with the flexibility to choose the option that best suits their needs and preferences for detecting leakage.

\subsection{Quick Fixes}
\label{sec:quick-fix}
\subsubsection{Overlap Leakage} It occurs when the training and test sets overlap; in other words, the two sets share data. Quick fixes for Overlap Leakage are associated with the sources of Overlap Leakage. Quick fixes may be applied to Overlap Leakage sources whose causes are \textit{SplitBeforeSample}, \ie splitting is performed before sampling. The plugin moves the existing call to a function with the keyword \say{split} above the nearest existing call to the function with the keyword \say{sample}. The plugin also adds a \say{TODO} message above the new call to \say{split} that instructs the user to check their code. The applied fix may not be syntactically or semantically correct, as the user must complete the fix. An example of an instance of Overlap Leakage (after the quick fix has been applied) is shown below in Listing \ref{Listing:overlap}.
\begin{lstlisting}[caption=A quick fix applied to Overlap Leakage., label=Listing:overlap, numbers=none, firstnumber = last, escapeinside={(*@}{@*)}]
    #TODO: Check the arguments provided to the call to split.
    X_train, X_test, y_train, y_test = train_test_split(
    X_new, y_new, test_size =0.2, random_state = 42)
    X_new, y_new =  (*@\colorbox{light-green}{SMOTE().fit\_resample (X,y)}@*)
       

\end{lstlisting}


\subsubsection{Multi-test Leakage} It occurs when the same test data is used in multiple evaluations. To apply a quick fix for Multi-test Leakage, the user hovers over one of the variables associated with a particular instance of Multi-test Leakage and selects \say{\textit{Use new test data for each evaluation}}. Each variable associated with the Multi-test Leakage will have a suffix of the form \say{\_\#” (\eg \_4) where the “\#} is one fewer than how many times that variable is reused. For example, if the last usage of the variable associated with an instance of Multi-test Leakage is renamed from \texttt{X\_test} to \texttt{X\_test\_4}, \texttt{X\_test} was used five times before the user applied the quick fix. The plugin also adds a \say{TODO} message above the newly renamed variable that instructs the user to load their test data into that variable. An example of a Multi-test Leakage instance (after the quick fix has been applied) is shown in Listing \ref{Listing:multitest}.

\begin{lstlisting}[caption=A quick fix applied to Multi-test Leakage., label=Listing:multitest, numbers=none, firstnumber = last, escapeinside={(*@}{@*)}]
    #TODO: Load the test data for the evaluation.
    lr_score = lr.score (X_test_0 , y_test)
       

\end{lstlisting}


\subsubsection{Preprocessing Leakage} It occurs when training and test data are transformed together. Quick fixes for Preprocessing Leakage are associated with the sources of Preprocessing Leakage. The plugin handles two distinct scenarios where Preprocessing Leakage occurs: (1) a split is performed after feature selection, and (2) no split is performed. To address the first case, the quick fix \textit{moves the split before the feature selection}. To address the second case, \textit{a new line of code must be added to split the data}. In essence, if a split call is already present, the quick fix simply moves that split call above the feature selection. If there is no split call present, the plugin adds a line containing \textsc{split()} above the feature selection. The plugin also adds a \say{TODO} message above the new call to \say{split} that instructs the user to check their code. It is up to the user to complete the code and ensure that it is correct. An example of a preprocessing leakage (after a quick fix has been applied) is shown in Listing \ref{Listing:preprocessing}. 

\begin{lstlisting}[caption=A quick fix applied to Preprocessing Leakage., label=Listing:preprocessing, numbers=none, firstnumber = last, escapeinside={(*@}{@*)}]
    #TODO: Check the arguments provided to the call to split.
    split()
    wordsVectorizer = (*@\colorbox{light-green}{CountVectorizer()}@*).fit(journalsFinal['text'])
    wordsVector = wordsVectorizer.transform(journalsFinal['text'])
       

\end{lstlisting}


It is worth noting that quick fixes simply provide a skeleton for a true fix. They are not substitutes for human judgement. The user needs to complete the fix and ensure the syntactical and semantical correctness of their code. A comprehensive fully automated solution is challenging, as it requires a deep conceptual understanding of the problem at hand. For example, when selecting features to use for model training, a developer might erroneously choose to use all features, rather than only a subset of features.  Consider, for example, a model that is intended to predict the yearly salary of a person based on a variety of factors. If a data point such as the monthly salary of the person appears in the training dataset, the model will simply use the monthly salary as an indicator of the yearly salary and ignore the other factors \cite{burkov2020machine}. A model trained on the wrong features, or more features than necessary, will not perform well in a production environment.



\section{Preliminary Evaluation}
\label{Section:Result}

We conducted a study to evaluate the effectiveness of \toolname in detecting Data Leakage and fixing them. Since \toolname is built on top of the leakage static analysis tool, we did not evaluate the detection accuracy of the tool independently. Instead, we considered the performance reported by Yang \etal \cite{yang2022data} to be valid for our purposes as well.

Since there is only one existing publicly available dataset \cite{yang2022data} that contains Data Leakage, we decided to use it for our own validation set. We started by selecting 31 Python files and manually analyzing the existing annotated sets and running them using our plugin. We invited 8 participants from Stevens
Institute of Technology to use any of these files when trying our plugin. All participants volunteered for the experiment, and informed consent was obtained.  The industrial experience of the
respondents ranged from 1 to 6 years, their ML
experience ranged from 1 to 6 years, their expertise
in programming ranged from 1 to 6 years, and their expertise
in Python ranged from 1 to 6 years. Before the experiment, the participants received an hour-long tutorial
on Data Leakage along with reference materials.


To evaluate the usefulness of our plugin, we have posted
a survey for users to take it optionally. The survey consisted of 15 questions that are divided
into 2 parts. The first part of the survey includes demographic questions about the participants. In the second part,
we ask about the (1) found Data Leakage type, (2)
level of satisfaction with the tool, (3)  additional information to be added to the tool window, (4) feasibility of implementing the quick fixes, and (5) overall impression of the tool and the proposed idea for improvement. As suggested by Kitchenham and Pfleeger \cite{kitchenham2008personal}, we constructed
the survey to use a 5-point ordered response scale (\say{Likert
scale}) question on the level of satisfaction with the tool,
5 open-ended questions on the tool experience and challenges, and 2 multiple choice questions on the type of Data Leakage found and the ability to see the tool window with all related information. All participants completed the survey. 

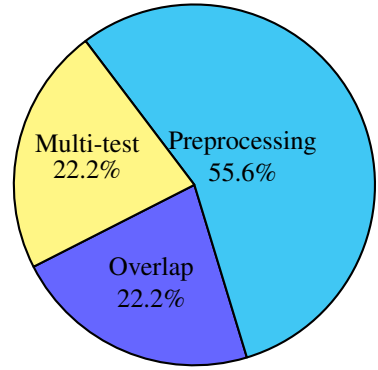
\begin{figure}
\centering 
\begin{tikzpicture}
\begin{scope}[scale=0.8]
\pie[rotate = 207,pos ={0,0},text=inside,no number]{22.2/Overlap\and22.2\%, 55.6/Preprocessing\and55.6\%, 22.2/Multi-test\and22.2\%}
\end{scope}
\end{tikzpicture}
\caption{Distribution of Data Leakage types selected by participants.} 
\label{fig:leakagetypess}
\vspace{-.6cm}
\end{figure}

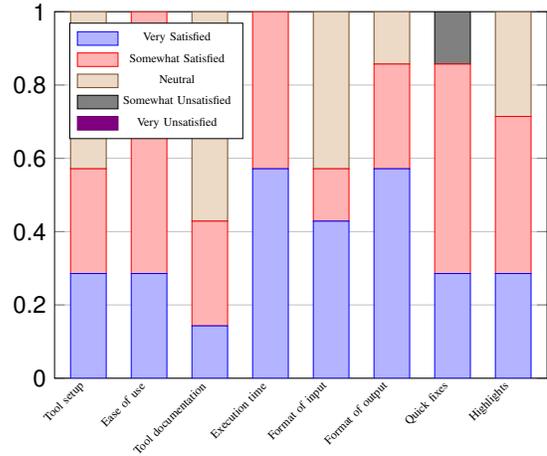
\begin{figure}
\centering
\begin{tikzpicture}
\begin{scope}[scale=0.9]
\begin{axis}[  width=\columnwidth, 
    height=7cm, 
    ybar stacked, 
    bar width=15pt, 
    xtick=data,
    symbolic x coords={Tool setup, Ease of use, Tool documentation, Execution time, Format of input, Format of output, Quick fixes, Highlights},
    xticklabel style={rotate=45, anchor=east, font=\tiny}, 
    ymajorgrids,
    ymin=0, ymax=1, 
    legend pos=north west,
    legend style={font=\tiny},
    enlarge x limits={abs=0.5cm}, 
    area legend
             ]

    \pgfplotstableset{
        create on use/total/.style={
            create col/expr={\thisrow{B} + \thisrow{C} + \thisrow{D}}
        }
    }

    \addplot table [x=Label, y=B_fraction]{\datatable};
    \addplot table [x=Label, y=C_fraction]{\datatable};
    \addplot table [x=Label, y=D_fraction]{\datatable};
    \addplot table [x=Label, y=E_fraction]{\datatable};
    \addplot table [x=Label, y=F_fraction]{\datatable};
     \legend{Very Satisfied, Somewhat Satisfied, Neutral, Somewhat Unsatisfied, Very Unsatisfied}
\end{axis}
\end{scope}
\end{tikzpicture}
\vspace{-.3cm}
 \caption{Participants’ satisfaction with various aspects of the \toolname tool.}
   \label{fig:usefulness}
\end{figure}

We report the results of the
users who have taken it in Figures \ref{fig:leakagetypess} and \ref{fig:usefulness}. Overall, as illustrated in Figure \ref{fig:leakagetypess}, the proportions of encountering different types of Data Leakage were similar for Overlap and Multi-test leakages. However, the most frequently observed was Preprocessing leakages, each accounting for 55.6\% of the total cases. This is expected as this type of leakage is more readily made in code than Multi-test. Multi-test leakage deals with one variable being used multiple times. A quick fix for this would be to make different variable names. Since the quick fix for the leakage is simply a name change, users who run into this problem probably just had simple naming mistakes and the fact that they need more variables.  Regarding user satisfaction, the survey results varied from `Very Satisfied', mainly with the tool setup, ease of use, execution time,  format of input/output, quick fixes and highlights, and mostly neutral opinions with the tool documentation. However, a user is less satisfied with the quick fixes.  The somewhat unsatisfaction can be explained by quick fixes that do not completely fix the code.  A comprehensive and fully automated correction is challenging, as it requires a deep conceptual
understanding of the problem.

\section{Implications}
\label{Section:Implication}


\noindent\textbf{Enhancing maintenance and evolution with \toolname.}
Our tool helps data scientists and ML model developers identify and fix instances of Data Leakage in their code.  Data Leakage is common in various types of notebooks, including those used for educational purposes \cite{yang2022data}. Data Leakage is often the result of human oversight, making it challenging for developers to spot in their own code. Regardless of the purpose of an ML model, Data Leakage can have significant consequences. In an educational setting, learners might adopt bad habits if they see examples of code with Data Leakage. In production environments, the impact can be more immediate, potentially causing a model to underperform in critical applications.

\noindent\textbf{Promoting adoption of coding best practices.} Our tool is designed to facilitate extensions and integration with other tools. This makes it a valuable resource for bridging the gap between research and practice in software maintenance and evolution. By simplifying deployment and use, \toolname can help practitioners access advanced tools more easily, promoting the adoption of best practices in the industry. The architecture of \toolname is designed to reduce the effort required for installation and usage, making cutting-edge research tools more accessible to practitioners.

\section{Related Work}
\label{Section:RelatedWork}

Kery \etal \cite{kery2018story}  
 interviewed 21 data scientists to study coding behaviors, and how they keep track
of variants they explored. Their findings stimulate the development of novel designs for interacting with notebook cells, facilitating tasks such as browsing history, debugging, and more, which could enhance the efficiency of literate programming in supporting data science endeavours.  Yang \etal \cite{yang2022data} proposed a static analysis approach to detect common forms of Data Leakage (\ie Overlap, Multi-test, and Preprocessing) in data science code. Bouke and Abdullah \cite{bouke2023empirical} investigated the impact of Data Leakage during data preprocessing on the reliability of machine learning based intrusion detection systems. The study employs six models on the datasets with and without Data Leakage to compare their performance. 
 Apicella \etal \cite{apicella2024don} discussed Data Leakage issues in the machine learning pipeline.  The authors classified Data Leakage in machine learning and examined how specific circumstances can spread throughout the machine learning process workflow. Drobnjakovi{\'c} \etal \cite{drobnjakovic2024abstract} developed a static analysis based on abstract interpretation to verify the absence of Data Leakage. Their approach was incorporated into the NBLyzer framework and was evaluated for its effectiveness and accuracy using 2111 Jupyter notebooks sourced from the Kaggle competition platform. Venkatesh \etal \cite{venkatesh2023enhancing} introduced HeaderGen, a tool that autonomously labels code cells with categorical markdown headers using a taxonomy for machine learning operations and classifies and displays function calls based on this classification. Wang \etal \cite{wang2022documentation} proposed Themisto toot to help write documentation for code cells by applying a deep learning-based approach
to generate documentation in natural language
and then recommending to the user whether to adopt it or
use it directly.

\section{Conclusion and Future Work}
\label{Section:Conclusion}

We develop \toolname, a PyCharm plugin that supports the detection and correction of Data Leakage. We conducted a preliminary assessment of the capabilities of the tool and plan to perform more empirical experiments as part of our future work. Additionally, it may be worth exploring the possibility of supporting other forms of Data Leakage. Further, the leakage analysis tool examines a variety of function calls from popular ML packages (\eg Keras, Sklearn). Our plugin provides plain language descriptions of leakage instances, identifies their sources and causes, and proposes fixes. In the future, we plan to support additional Data Leakage causes as the plain language description of a leakage instance (and its cause) depends on the function calls involved in that leakage instance.  

\section{Acknowledgments}
\label{sec:ack}
We would like to thank the authors of the leakage static analysis tool \cite{yang2022data} for publicly providing it.


\bibliographystyle{abbrv}

\bibliography{IEEEabrv,sample-base}

\end{document}